\documentclass[aps,prl,twocolumn,preprintnumbers,showpacs]{revtex4}        
\usepackage{graphicx}
\usepackage{amsfonts}
\usepackage{amsmath}
\usepackage{amssymb}
\usepackage{amsfonts}

\begin{document}
\title{Effective one-dimensional description of confined diffusion biased
by a transverse gravitational force}

\author{Pavol Kalinay}

\affiliation{Institute of Physics, Slovak Academy of
Sciences, D\'ubravska cesta 9, 84511, Bratislava, Slovakia}

\date{}

\begin{abstract}
Diffusion of point-like non interacting particles in a two-dimensional (2D)
channel of varying cross section is considered. The particles are biased
by a constant force in the transverse direction. We apply our recurrence
mapping procedure, which enables us to derive an effective one-dimensional
(1D) evolution equation, governing the 1D density of the particles in the
channel. In the limit of stationary flow, we arrive at an extended
Fick-Jacobs equation, corrected by an effective diffusion coefficient
$D(x)$, depending on the longitudinal coordinate $x$. Our result is an
approximate formula for $D(x)$, involving also influence of the transverse
force. Our calculations are verified on the stationary diffusion in a linear
cone, which is exactly solvable.
\end{abstract}

\pacs{05.40.Jc, 87.10.Ed}

\maketitle

\renewcommand{\theequation}{1.\arabic{equation}}
\setcounter{equation}{0}

\section{I. Introduction}

A point-like particle diffusing in a two or three dimensional (2D,3D)
channel of varying cross section became an archetypal model describing
transport through nano channels, pores or along fibers in biological
systems, as well as passing of large molecules through membranes.
Analytic studies of such models usually require further simplifications,
namely the dimensional reduction to a purely one-dimensional (1D)
system described by an effective 1D evolution equation, governing
the linear (1D) density $p(x,t)$, depending on time $t$ and the
longitudinal coordinate $x$. On the other hand, any simplification
should retain all important features of the original full dimensional
model. After the dimensional reduction, they are reflected in the
structure of the effective equation.

The Fick-Jacobs (FJ) equation \cite{FJ},
\begin{equation} \label{1.1}
\partial_t p(x,t)= D_0 \partial_x A(x)\partial_x \frac{p(x,t)}{A(x)},
\end{equation}
can serve as the simplest example of such an effective 1D equation for
diffusion in a channel with reflecting walls; $A(x)$ denotes the
cross section area for 3D or the width for 2D channels at some $x$,
and $D_0$ is the diffusion constant. This equation maintains only the
mass conservation along the channel with varying $A(x)$. Introducing
a spatially dependent effective diffusion coefficient $D(x)$ in the
effective equation \cite{20,RR},
\begin{equation} \label{1.2}
\partial_t p(x,t)=\partial_x A(x)D(x)\partial_x \frac{p(x,t)}{A(x)},
\end{equation}
enables us also to respect boundary conditions (BC) and the local 
mass conservation at a point $(x,{\bf y})$ of the full dimensional
problem in the case of the stationary flow, i.e. when the net flux
$J(x,t)$ flowing through the channel is constant ({\bf y} denotes the
transverse coordinates). For nonstationary processes, one should
replace the function $D(x)$ by an operator $\hat D(x)$, containing also
the spatial derivatives $\partial_x^n$, $n=1,2,...$, but in practice,
in most cases the asymptotic behavior of the processes is studied and the FJ
equation extended only by the function $D(x)$, Eq. (\ref{1.2}),
represents a significant improvement of the standard FJ approximation
(\ref{1.1}).

Of course, it is necessary to find a way how to fix the function $D(x)$.
Based on a fenomenological argumentation, Reguera and Rub\'i \cite{RR}
suggested a function
\begin{equation} \label{1.3}
D(x)=D_0\big[1+h'^2(x)\big]^{-\eta}\ ,
\end{equation}
where $\eta=1/3$ or $1/2$ for 2D or 3D channels with axial symmetry,
respectively, and $h(x)$ denotes half width or radius of the channel.
Later Kalinay and Percus \cite{map}-\cite{eff} showed, that 
this function can be found as a series expansion in a small parameter
$\epsilon$, representing the ratio of the longitudinal and the transverse
diffusion constant, $\epsilon=D_0/D_y$. The anisotropy of the diffusion
constant, imposed artificially, causes separation of the modes quickly
decaying in the transverse direction from much slower longitudinal ones
and formally, it enables us to find a recurrence scheme, generating
systematically higher order corrections to the FJ equation (\ref{1.1}),
giving the expansion of the function $D(x)$ in $\epsilon$ in the
stationary state. For 2D channels, bounded by $y=h(x)$ and the $x$ axis,
we get
\begin{eqnarray} \label{1.4}
D(x)&=&D_0\Big(1-\frac{\epsilon}{3}h'^2+\frac{\epsilon^2}{45}h'\cr
&&\times\big[9h'^3+hh'h''-h^2h^{(3)}\big]-...\Big) .
\end{eqnarray}
If $h''(x)$ and the higher derivatives are neglected, the result is
\begin{equation}\label{1.5}
D(x)\simeq D_0\Big(1-\frac{\epsilon}{3}h'^2+\frac{\epsilon^2}{5}h'^4-
...\Big)=D_0\frac{\arctan\sqrt{\epsilon}h'}{\sqrt{\epsilon}h'},
\end{equation}
a function differing from (\ref{1.3}) by less than 1\% for moderate
slopes, $|h'|<1$, for an isotropic diffusion, $\epsilon=1$. For 3D
symmetric channels, the same treatment results in the formula
(\ref{1.3}) and $\eta=1/2$. All formulas exhibited good agreement
with numerical tests \cite{Berez} for $|h'|<1$; steeper slopes
require to take also the higher derivatives of $h(x)$ into
account \cite{appr}, or to avoid the expansion for specific geometries
at all \cite{Ber}-\cite{Dagdug}.

Let us remark that introducing anisotropy of the diffusion constant is
equivalent to rescaling of all transverse lengths (i.e. the coordinate
$y$ and the half width $h(x)$ in the 2D channels) by $\sqrt{\epsilon}$
\cite{exact}. This scaling together with a similar recurrence procedure
were used to calculate corrections to the mean velocity and the
dispersivity of the particles \cite{YD1,YD2} within the macrotransport
theory, as well as for re-derivation of $D(x)$ in Eq. (\ref{1.5})
\cite{MaH}.

Recent studies \cite{forced, soft} showed that the same strategy could
be used also for the dimensional reduction of the Smoluchowski equation,
\begin{eqnarray}\label{1.6}
\partial_t\rho(x,{\bf y},t)&=&\Big[D_0\partial_xe^{-U(x,{\bf y})/k_BT}
\partial_xe^{U(x,{\bf y})/k_BT}+\hskip0.4in\cr
&&\hskip-0.5in D_y\nabla_y\cdot e^{-U(x,{\bf y})/k_BT}\nabla_y
e^{U(x,{\bf y})/k_BT}\Big]\rho(x,{\bf y},t),
\end{eqnarray}
i.e. for mapping of the diffusion in an external field $U(x,{\bf y})$;
$\rho(x,{\bf y},t)$ is the 2D or 3D density, $\nabla_y$ denotes the
gradient in the transverse directions and $T$ is the temperature.
The ratio $\epsilon=D_0/D_y$, remains a good small parameter also for
the biased diffusion, enabling us again to construct a recurrence
procedure generating systematically corrections to an equivalent of the
FJ equation. This extension allows us to apply the dimensional reduction
to a much broader class of problems interesting in chemical physics.
Considering a force parallel to the $x$ axis of the channel, we showed
\cite{forced} how the entropic potential is added to the real (energetic)
one in the dimensionally reduced dynamics, which can be useful for studying
e.g. the Brownian pumps \cite{Liu}. Mapping of Eq. (\ref{1.6}) for a
potential depending on the transverse coordinates \cite{soft} can be
used to get the reduced dynamics in a channel with soft walls.

\begin{figure}
\includegraphics[scale=0.33]{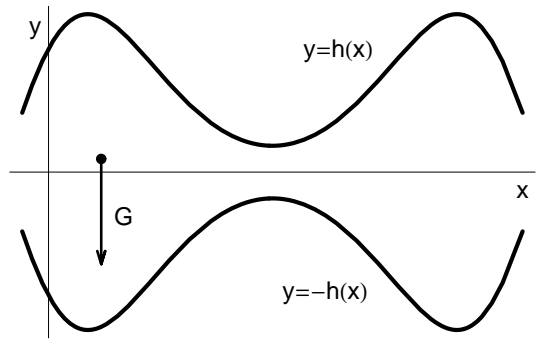}
\caption{A sketch of the considered model: the channel is bounded by
hard walls at $y=h(x)$ and $-h(x)$. Diffusing particles are biased by
a constant gravitational force $G$}
\end{figure}
In the present paper, we study diffusion in a 2D symmetric channel,
bounded by smooth functions $y=h(x)$ and $-h(x)$, with hard and
reflective walls (Fig.1). The particles are biased in the transverse 
direction by a constant gravitational force $G$; the potential
$U(x,y)=Gy$ in Eq. (\ref{1.6}). This model was investigated mainly
in connection with the stochastic resonance during the last years
\cite{SR}-\cite{xpx}. An important feature of this model is an interplay
between the gravitational force, holding particles in the potential wells
in the wider parts of the channel, and the thermal motion, enabling the
particles to diffuse into the neighboring compartments over the
potential barriers formed by the narrowings of the channel.
An oscillating force applied along the channel can help this hopping
very effectively at a specific resonance frequency, depending on 
the force $G$, the temperature $T$ and the geometry of the channel.

We focus our attention on the competition between the gravitational
potential and the "entropic" potential in diffusion through such
channels. Our mapping procedure allows us to quantify their
contributions to the net flow of particles in an elegant way: in the
form of an effective 1D equation, involving both effects in its structure.
The previous analyses used the 1D description, too, but governed by
an equivalent of the FJ equation (\ref{1.1}). Comparison of this
theory with the Brownian simulations \cite{Das,Ray} indicates that this
approximation may be not satisfactory especially in the region of our
interest, when both effects become comparable.

In the following section, we present the rigorous dimensional
reduction of this model onto the longitudinal coordinate. We
show how to merge the mapping of diffusion in a transverse field
\cite{soft} with the presence of the reflecting hard walls.
The result of our mapping is an equation of the type (\ref{1.2})
in the limit of the stationary flow, with $D(x)$ expanded to the
first few orders in $\epsilon$. In Section III, we suggest
and justify an interpolation formula for $D(x)$ based on 
results of the mapping procedure in the "linear approximation",
when $h''(x)$ and higher derivatives in the expansion of $D(x)$
are neglected. Our formula is verified by an exactly solvable model,
the stationary diffusion of particles in a linear cone.

\renewcommand{\theequation}{2.\arabic{equation}}
\setcounter{equation}{0}

\section{II. Mapping procedure}

We follow the procedure developed for the diffusion \cite{map,eff}
and the biased diffusion \cite{forced,soft}, based on introducing
a small parameter $\epsilon$ into the 2D Smoluchowski equation
(\ref{1.6}). We define it as a parameter of anisotropy of the diffusion
constant, $D_y=D_0/\epsilon$, but it can be also imposed by scaling of
the transverse lengths \cite{MaH}, $y\rightarrow\sqrt{\epsilon}y$
and the inverse scaling of the force $G\rightarrow G/\sqrt{\epsilon}$.
Anyway, we get the equation
\begin{equation}\label{2.1}
\partial_t\rho(x,y,t)=\partial_x^2\rho(x,y,t)+\frac{1}{\epsilon}
\partial_y e^{-gy}\partial_y e^{gy}\rho(x,y,t)
\end{equation}
from Eq. (\ref{1.6}) for the model of our interest; we rescaled time $t$
by the diffusion constant $D_0$, $D_0t\rightarrow t$, and $g=G/k_BT$.

The small parameter $\epsilon$ enables us to carry out two important
steps of the mapping procedure. First, we can find readily the equivalent
of the FJ equation in the limit $\epsilon\rightarrow 0$, and second,
it becomes a parameter controlling the perturbation expansion of any
quantity describing diffusion in the channel: the 2D density $\rho(x,y,t)$
and the flux density ${\bf j}(x,y,t)$ \cite{MaH}, 1D density $p(x,t)$,
or any mean value, like the mean velocity or dispersivity \cite{YD1,
YD2}, representing a sequence of corrections to the zero-th order (FJ)
solution.

Before the mapping, we have to supplement boundary conditions (BC).
The Smoluchowski equation (\ref{2.1}) represents the mass conservation
law, so the components of the flux density ${\bf j}$ are
\begin{eqnarray} \label{2.2}
j_x(x,y,t)&=&-\partial_x\rho(x,y,t)\ ,\cr
j_y(x,y,t)&=&-\frac{1}{\epsilon}e^{-gy}\partial_ye^{gy}\rho(x,y,t)\ .
\end{eqnarray}
No flux through the reflecting hard walls requires to have the vector
{\bf j} at the boundaries parallel to them, so we get
\begin{equation}\label{2.3}
e^{gy}\partial_ye^{-gy}\rho(x,y,t)=\pm\epsilon h'(x)\partial_x
\rho(x,y,t)\Big|_{y=\pm h(x)}
\end{equation}
at the upper and the lower boundary $y=\pm h(x)$. BC at the ends of the
channel are arbitrary, they do not enter the mapping procedure in our
formulation. We can consider the channel as infinite.

The mapping procedure reduces the 2D Smoluchowski equation (\ref{2.1})
governing the 2D density $\rho(x,y,t)$ to some 1D equation governing
the 1D density $p(x,t)$, defined as
\begin{equation}\label{2.4}
p(x,t)=\int_{-h(x)}^{h(x)}\rho(x,y,t)dy.
\end{equation}
Thus the first step of the mapping is integration of Eq. (\ref{2.1})
over the cross section. Applying the definition (\ref{2.4}) on the left
hand side, we arrive at
\begin{eqnarray}\label{2.5}
\partial_tp(x,t)&=&\int_{-h(x)}^{h(x)}\partial_x^2\rho(x,y,t)dy\cr
&&\hskip0.5in+\frac{1}{\epsilon}\left[e^{-gy}\partial_ye^{gy}
\rho(x,y,t)\right]_{-h(x)}^{h(x)}\cr
&=&\partial_x\int_{-h(x)}^{h(x)}\partial_x\rho(x,y,t)dy
\end{eqnarray}
after integrating by parts and using BC (\ref{2.3}). 

Our goal is also to express the right hand side of Eq. (\ref{2.5})
in terms of $p(x,t)$ instead of $\rho(x,y,t)$. This task is easy
to complete in the limit $\epsilon\rightarrow 0$. For an infinitesimally
small $\epsilon$, the transverse diffusion constant $D_y$ becomes almost
infinite and the 2D density $\rho_0$ is equilibrated in the transverse
direction almost immediately after any change in the $x$ direction.
So we can write
\begin{equation}\label{2.6}
\rho_0(x,y,t)=\frac{1}{A(x)}e^{-gy}p(x,y),
\end{equation}
where $A(x)$ provides normalization of $\rho_0$. If substituted in
the condition (\ref{2.4}), we have to obtain an identity. Hence
\begin{equation}\label{2.7}
A(x)=\int_{-h(x)}^{h(x)}e^{-gy}dy=\frac{2}{g}\sinh{[gh(x)]}.
\end{equation}
If we use the formula (\ref{2.6}) for the 2D density in Eq. (\ref{2.5}),
we find
\begin{eqnarray}\label{2.8}
\partial_tp(x,t)&=&\partial_x\left[\int_{-h(x)}^{h(x)}e^{-gy}dy\right]
\partial_x\frac{p(x,t)}{A(x)}\cr &=&\partial_xA(x)\partial_x
\frac{p(x,t)}{A(x)},
\end{eqnarray}
which is (an equivalent of) the FJ equation (\ref{1.1}). Let us stress
that $A(x)$ is not the width of the channel here, but the integral
(\ref{2.7}). On the other hand, for $g\rightarrow 0$, $A(x)$
becomes $2h(x)$.

For $\epsilon>0$, the transverse diffusion constant $D_y$ is finite
and the local equilibrium in the $y$ direction is disturbed by the
flux flowing along the curved boundaries at a given $x$. So the formula
(\ref{2.6}) cannot be used for the 2D density, $\rho_0(x,y,t)$ does not
satisfy the Smoluchowski equation (\ref{2.1}). The small parameter
$\epsilon$ enables us to look for deviations of the real 2D density
$\rho(x,y,t)$ from the equilibrated density $\rho_0$ (\ref{2.6}) in the
form of a sequence of corrections, as an expansion in powers of $\epsilon$.

Instead of expanding some specific density $\rho(x,y,t)$ in $\epsilon$,
we prefer a more general method. We look for the expansion of a wide class
of solutions of the 2D problem; we expand all $\rho(x,y,t)$, which can
be generated from any 1D solution $p(x,t)$ of the searched 1D equation
by the backward mapping onto the space of solutions of the 2D problem.
Formally, these 2D densities can be expressed by the formula
\begin{equation}\label{2.9}
\rho(x,y,t)=e^{-gy}\hat\omega(x,y,\partial_x)\frac{p(x,t)}{A(x)},
\end{equation}
where $\hat\omega$ represents an operator of the backward mapping.
It acts on a wide class of $p(x,t)$, so the dependence of $p(x,t)$
on $\epsilon$ is not important, we take it as an unfixed function.
Instead, we look for expansion of the operator $\hat\omega$ in
$\epsilon$, so
\begin{equation}\label{2.10}
\rho(x,y,t)=e^{-gy}\sum_{n=0}^{\infty}\epsilon^n\hat\omega_n
(x,y,\partial_x)\frac{p(x,t)}{A(x)}.
\end{equation}
We know already the zero-th order term, $\hat\omega_0=1$, which
gives the equilibrated solutions $\rho_0$ (\ref{2.6}) in the limit
$\epsilon\rightarrow 0$.

Supposing $\rho(x,y,t)$ of the form (\ref{2.10}), we can formally
complete the construction of the 1D evolution equation. Applying this
expression in Eq. (\ref{2.5}), we find
\begin{eqnarray}\label{2.11}
\partial_t p(x,t)&=&\partial_x \int_{-h(x)}^{h(x)}dye^{-gy}\partial_x
\sum_{n=0}^{\infty}\epsilon^n
\hat\omega_n(x,y,\partial_x)\frac{p(x,t)}{A(x)}\cr
&=&\partial_x A(x)\Big[1-\epsilon\hat Z(x,\partial_x)\Big]
\partial_x\frac{p(x,t)}{A(x)}.
\end{eqnarray}
Here we already applied $\hat\omega_0=1$ and introduced an operator
$\hat Z$, correcting the FJ equation (\ref{2.8}). It can also be
expanded in $\epsilon$,
\begin{eqnarray}\label{2.12}
\epsilon\hat Z(x,\partial_x)=\sum_{k=1}^{\infty}\epsilon^k\hat Z_k
(x,\partial_x);\hskip1in\cr
\hskip-0.2in\hat Z_k(x,\partial_x)\partial_x=-\frac{1}{A(x)}
\int_{-h(x)}^{h(x)}dy e^{-gy}\partial_x\hat\omega_k(x,y,\partial_x).
\end{eqnarray}

Finally, the 2D density (\ref{2.10}) has to be a solution of the
original Smoluchowski equation (\ref{2.1}),
\begin{equation}\label{2.13}
\sum_{j=0}^{\infty}\epsilon^j\Big[e^{-gy}(\partial_t-\partial_x^2)-
{1\over\epsilon}\partial_ye^{-gy}\partial_y\Big]\hat\omega_j(x,y,
\partial_x){p(x,t)\over A(x)}=0,
\end{equation}
which generates a recurrence relation fixing the operators
$\hat\omega_n$. Because we suppose that these operators act only on
the spatial coordinates, the time derivative commutes with them and
for $\partial_tp(x,t)$, we use the equation (\ref{2.11}). Collecting
the terms at the same powers of $\epsilon$, we find
\begin{eqnarray}\label{2.14}
\partial_ye^{-gy}\partial_y\hat\omega_{n+1}(x,y,\partial_x)=
-e^{-gy}\Big[\partial_x^2\hat\omega_n(x,y,\partial_x)\hskip0.3in\cr
+\sum_{k=0}^n\hat\omega_{n-k}(x,y,\partial_x){1\over A(x)}
\partial_x A(x)\hat Z_k(x,\partial_x)\partial_x\Big];\hskip0.2in
\end{eqnarray}
we take $\hat Z_0(x,\partial_x)=-1$ in this formula. After double
integration, we obtain $\hat\omega_{n+1}$, giving the $n+1$-st correction
to the 2D density, if applied on some 1D solution $p(x,t)$. Two
integration constants have to be fixed (they are also operators, but
independent of $y$). The first one provides satisfaction of the BC
(\ref{2.3}). Using the formula (\ref{2.10}) in Eq. (\ref{2.3}) and
comparing the terms at the same powers of $\epsilon$, we get the condition
\begin{equation}\label{2.15}
\partial_y\hat\omega_{n+1}(x,y,\partial_x)=\pm h'(x)\partial_x
\hat\omega_n(x,y,\partial_x)\Big|_{y=\pm h(x)}.
\end{equation}
If the integration constant is fixed at one boundary, the BC at the
opposite boundary is automatically satisfied. The second integration
constant helps to satisfy the normalization condition; applying
the formula (\ref{2.10}) in the definition (\ref{2.4}) has to give
identity in any order of $\epsilon$, hence
\begin{equation} \label{2.16}
\int_{-h(x)}^{h(x)}dye^{-gy}\hat\omega_{n}(x,y,\partial_x)=0
\end{equation}
for $n>0$. By this condition, we keep the same number of particles
in both, 1D and 2D descriptions, independently of how the densities
$\rho$ or $p$ are normalized.

The recurrence procedure starts from the FJ approximation, $\hat\omega_0
=1$ and $\hat Z_0=-1$. Having expressed $\hat\omega_{n+1}$ for some $n$,
the next order correction operator $\hat Z_{n+1}$ is calculated according
to the formula (\ref{2.12}). Calculation of the first order correction
and other details are given in the Appendix A. We show here only the results,
\begin{equation}\label{2.17}
\hat\omega_1=\frac{h'}{g}\bigg[\frac{e^{gy}+(1-gy)\cosh{gh}}{\sinh{gh}}
-gh\Big(1+\frac{2}{\sinh^2gh}\Big)\bigg]\partial_x
\end{equation}
and the corresponding
\begin{equation}\label{2.18}
\hat Z_1=\frac{h'^2}{\sinh^2gh}\Big[1+\cosh^2gh-2gh\coth{gh}\Big].
\end{equation}
One can check that in the limit $g\rightarrow 0$, we obtain $\hat Z_1
\rightarrow h^2/3$, known for the diffusion alone \cite{20,eff}.
The higher order operators $\hat Z_n$ starting from $n=2$ also contain
the spatial derivatives $\partial_x$, what makes the equation
(\ref{2.11}) too difficult for direct use in practice. Alike the
diffusion alone \cite{eff}, this equation can be simplified by
replacing the correction operator $1-\epsilon\hat Z$ by the function
$D(x)$ in the limit of the stationary state, when the net flux changes
very slowly.

In that case, Eq. (\ref{2.11}) is replaced by an equation of
the form (\ref{1.2}), where $A(x)$ is given by the formula (\ref{2.7})
and $D(x)$ has to be fixed. Thus we have two different expressions for
the net flux,
\begin{equation}\label{2.19}
J(x,t)=-A(x)\left[1-\epsilon\hat Z(x,\partial_x)\right]
\partial_x\frac{p(x,t)}{A(x)}
\end{equation}
and
\begin{equation}\label{2.20}
J(x,t)=-A(x)D(x)\partial_x\frac{p(x,t)}{A(x)},
\end{equation}
coming from Eqs. (\ref{2.11}) and (\ref{1.2}), respectively, as both
equations represent the 1D mass conservation law. In the stationary
state, $J(x,t)=J$ is constant in time and space and $\partial_x
[p(x,t)/A(x)]/J=-1/A(x)D(x)$ depends only on geometry and the parameters
of the model for any stationary solution $p(x,t)=p(x)$. Then the formula
(\ref{2.19}) describes the same flux $J$ only if
\begin{equation}\label{2.21}
{1\over D(x)}=A(x)\Big[1-\epsilon\hat Z(x,\partial_x)\Big]^{-1}
{1\over A(x)}\ ,
\end{equation}
which fixes the effective diffusion coefficient $D(x)$ unambiguously
for $\hat Z$ obtained from the mapping procedure. If the expansion
of $\hat Z$ in $\epsilon$ (\ref{2.12}) is used in Eq. (\ref{2.21}),
the result is an $\epsilon$-expansion of $D(x)$,
\begin{eqnarray}\label{2.22}
D(x)&=&1-\frac{\epsilon h'^2}{\sinh^2gh}\Big[1+\cosh^2gh-2gh\coth{gh}
\Big]\cr&&+\frac{\epsilon^2h'^4}{\sinh^6gh}\Big[\sinh^4gh\cosh^2gh-
\frac{gh}{2}\sinh(2gh)\cr &&\times\left(17\sinh^2gh+36\right)+(gh)^2
\big(7\sinh^4gh\cr && +40\sinh^2gh+36\big)\Big]+O(\epsilon^3)+O(h'');
\end{eqnarray}
the second and higher derivatives of $h(x)$ are already neglected in
this formula.

\renewcommand{\theequation}{3.\arabic{equation}}
\setcounter{equation}{0}

\section{III. Interpolation formula for D(${\bf x}$)}

Even in the "linear approximation", which neglects all but the first
derivative of $h(x)$ in the expansion of $D(x)$, the resulting formula
(\ref{2.22}) is much more complicated than the similar one valid for
the diffusion alone, Eq. (\ref{1.5}). Also it is difficult to sum
directly the series in $\epsilon h'^2$ up to infinity, to find a formula
for $D(x)$ in a closed form, usable in practice. In this section, we
propose an {\it ad hoc} formula, justify its validity and test it on an
exactly solvable model.

The expansion (\ref{2.22}) simplifies considerably in two limits:
for $g\rightarrow 0$ and $gh(x)\rightarrow\infty$. The first case
corresponds to the unbiased diffusion, the coefficients at
$(\epsilon h'^2)^n$ approach $(-1)^n/(2n+1)$ and the series is
summable, giving the result (\ref{1.5}). In a strong gravitational field, 
the limit of the coefficients is $(-1)^n$, so
\begin{equation}\label{3.1}
D(x)\rightarrow 1-\epsilon h'^2+\epsilon^2h'^4-...=
\frac{1}{1+\epsilon h'^2};
\end{equation}
the proof is given in the Appendix A.

\begin{figure}
\includegraphics[scale=0.4]{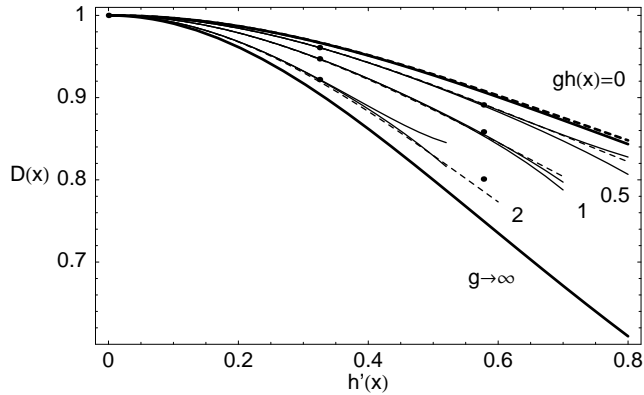}
\caption{The coefficient $D(x)$ depending on the local slope $h'(x)$ and
the values of $gh(x)=0,0.5,1,2$ and infinity at some point $x$. The thick
lines depict the limits $g\rightarrow 0$ and $\infty$. The dashed lines
correspond to the interpolation formula (\ref{3.2}) with the exponent
(\ref{3.3}). The adjacent thin full lines describe the truncated expansion
(\ref{2.22}) up to the 3-rd order (the lower lines) and the 4-th order
(the upper lines). The dots depict the data gained from the exactly solvable
model, a linear cone with $h'(x)=\tan(\pi/10)\simeq 0.325$ and $\tan(\pi/6)=
1/\sqrt{3}$.}
\end{figure}
Finally, we recall that the formula (\ref{1.3}) differs only slightly
from the exact result, Eq. (\ref{1.5}), at moderate slopes of the walls,
$|h'(x)|<1$. Then it seems reasonable to suggest this formula also for
the region of intermediate $g$, but with the exponent $-\eta$ depending
on $gh(x)$,
\begin{equation}\label{3.2}
D(x)\simeq D_0[1+\epsilon h'^2(x)]^{-\eta[gh(x)]}\ .
\end{equation}
For the choice
\begin{equation}\label{3.3}
\eta[gh(x)]=\frac{1}{\sinh^2gh}\Big[1+\cosh^2gh-2gh\coth{gh}\Big]\ ,
\end{equation}
we recover correctly the first order term of $D(x)$ in Eq. (\ref{2.22}).
Then in strong fields, the formula (\ref{3.2}) approaches the exact
limit (\ref{3.1}) and for $g\rightarrow 0$, we get the function of
Reguera and Rub\'i, Eq.(\ref{1.3}).

We compare first our interpolation formula with the true expansion
of $D(x)$ (\ref{2.22}) calculated up to the 4-th order in $\epsilon$.
The plots of $D(x)$ versus slope of the boundaries $h'(x)$ are depicted
in Fig. 2. The thick lines describe the limits, $g\rightarrow 0$ and
$\infty$. The dashed lines plot the formula (\ref{3.2}) for three
intermediate values of $g=0.5$, $1$ and $2$. These data are compared
with the truncated series (\ref{2.22}), the adjacent thin lines include
the corrections up to the 3-rd order (the lower lines), and the 4-th
order (the upper lines). In the region of fast convergence of the
series (\ref{2.22}), where the lines of the 3-rd and the 4-th order
formulas almost coincide, the difference between the true and the
interpolated values is comparable with the difference between the formulas
(\ref{1.5}) and (\ref{1.3}) (the thick dashed line).

Unfortunately, the radius of convergence of this series is finite and
decreasing with growing $g$. So we test our interpolation formula
on an exactly solvable model.

Tests of such theories are often based on calculation of the
net flux $J$ flowing through an exactly solvable structure. The flux
calculated from the exact 2D density $\rho(x,y)$,
\begin{equation}\label{3.4}
J=\int_{-h(x)}^{h(x)}j_x(x,y)dy=-\int_{-h(x)}^{h(x)}\partial_x\rho(x,y)dy
\end{equation}
in the stationary regime, is compared with the corresponding flux
according to Eq. (\ref{2.20}) with $D(x)$ derived within the tested
theory. We modify this method: for a given exact solution $\rho(x,y)$,
we calculate the flux $J$ (\ref{3.4}), the 1D density $p(x)$ (\ref{2.4})
and the corresponding $D(x)$,
\begin{equation}\label{3.5}
D(x)=-\frac{J}{A(x)}\left(\partial_x\frac{p(x)}{A(x)}\right)^{-1}
\end{equation}
from Eq. (\ref{2.20}), which is compared with $D(x)$ coming from the
theory, Eq. (\ref{3.2}) or (\ref{2.22}) in our case.

\begin{figure}
\includegraphics[scale=0.3]{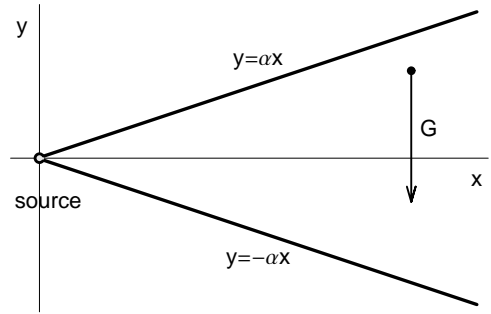}
\caption{Linear cone with a constant transverse force $G$.}
\end{figure}
Our exactly solvable model is a stationary flow through a linear
cone, bounded by $y=\pm\alpha x$, see Fig. 3. The particles are
emitted from a point-like source at the origin of the coordinate
system and collected at an absorbing boundary placed far from the
positions $x$ of our interest. Let us notice that the expansion
(\ref{2.22}) of $D(x)$ summed up to infinity describes our model
exactly; $h'(x)=\alpha$ is constant and its derivatives are zero.

In the Appendix B, we show that the 2D density expressed in the form
of an integral in the complex plane
\begin{eqnarray}\label{3.6}
\rho(x,y)&=&e^{-gy/2}\int_0^{i\pi/2+\infty}e^{-(g\sqrt{x^2+y^2}/2)
\cosh(z-i\pi/2)}\ \ \ \cr &&\times\Big[f(z+i\phi)+f(z-i\phi-i\pi)
\Big]dz +c.c.,
\end{eqnarray}
$\phi=\arctan{y/x}$ and $f(w)=\coth(mw/2)\tanh(w/2)$, $m=3,5,7,...$,
represents a stationary solution of the Smoluchowski equation
(\ref{2.1}) with BC (\ref{2.3}), $\epsilon=1$, $D_0=1$ and $h(x)=
\alpha x$ for specific values of the slope $\alpha=\tan{\phi_0}$;
$\phi_0=\pi/2m=\pi/6,\pi/10,..$. The integration from $0$ to $i\pi/2+
\infty$ (and to $-i\pi/2+\infty$ in the complex conjugated expression)
is carried out along any path avoiding the poles of the integrand on
the imaginary axis from the right side, see Fig. 7 in the Appendix B.
\begin{figure}
\begin{tabular}{c c}
\includegraphics[scale=0.3]{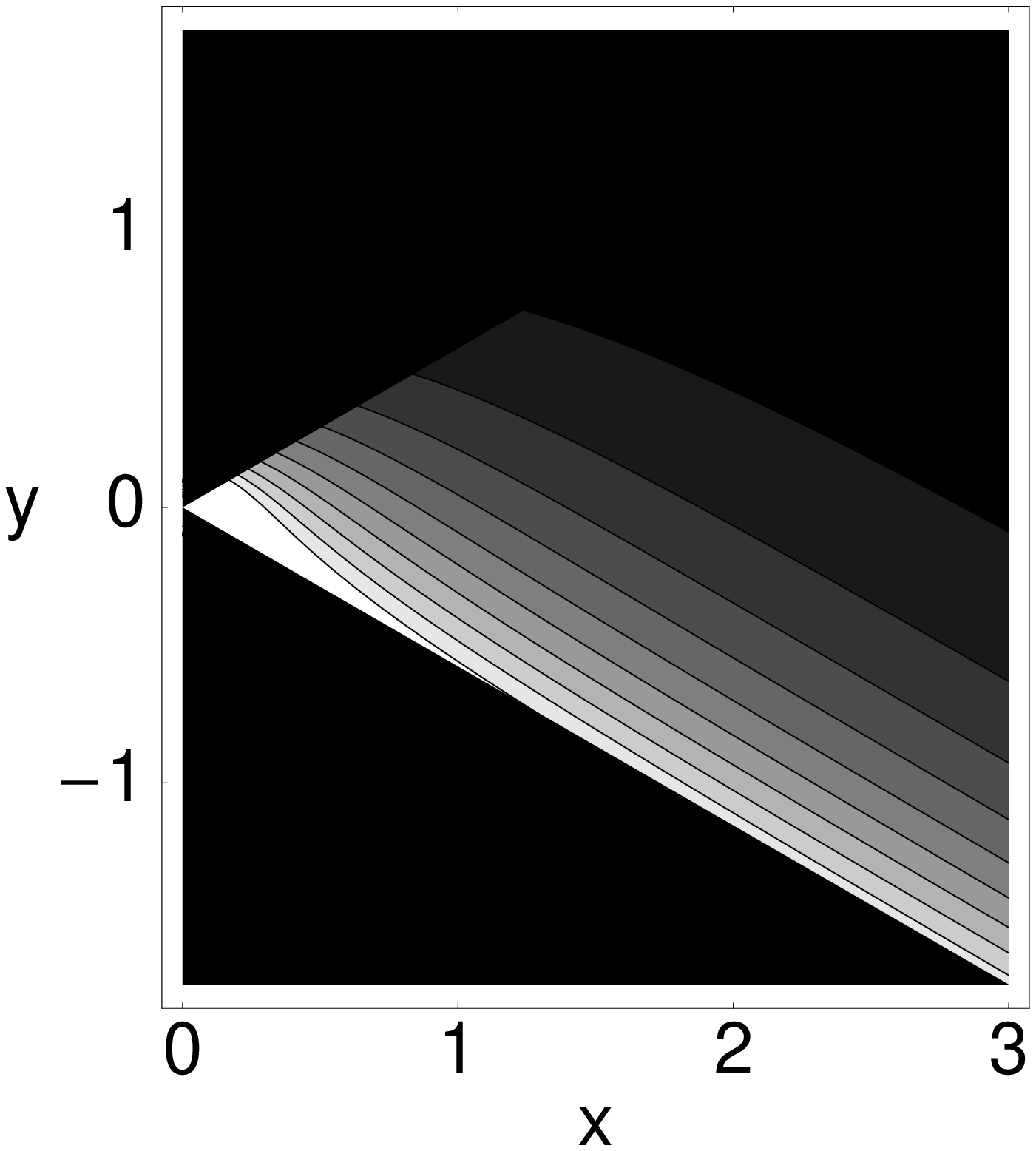}&
\includegraphics[scale=0.3]{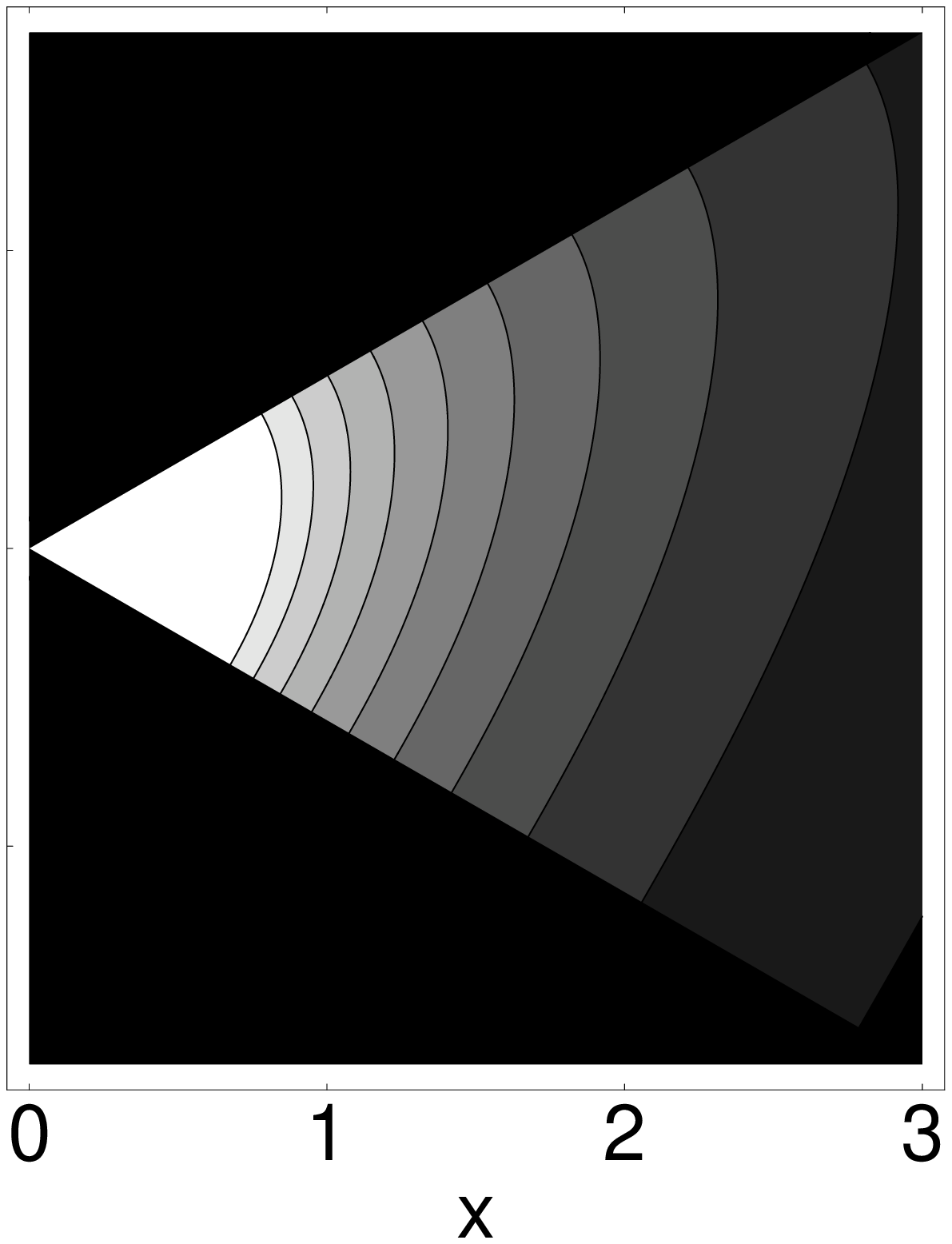}
\end{tabular}
\caption{Contour plot of the 2D density $\rho(x,y)$ (the left panel)
and $\bar\rho(x,y)=e^{gy}\rho(x,y)$ (the right panel) according to Eq.
(\ref{3.6}) in a channel bounded by $y=\pm x/\sqrt{3}$ and $g=2$.}
\end{figure}
The contour plots of the density $\rho$ (\ref{3.6}) and the
corresponding $\bar\rho(x,y)=e^{gy}\rho(x,y)$ are shown in Fig. 4.
According to Eq. (\ref{2.2}), the gradient of $\bar\rho(x,y)$
is proportional to the flux density, so one can check visually
on the right panel that the no flux BC are satisfied on both
boundaries.

For testing the interpolation formula (\ref{3.2}), we use the channels
with $\alpha=\tan(\pi/6)=1/\sqrt{3}$ and $\tan(\pi/10)\simeq 0.325$.
The values of $p(x)$ and $J$ in Eq. (\ref{3.6}) were integrated
numerically; the calculation of $J$ serves as a test of the numerical
method, since $J$ does not depend on $x$. A choice of the force $g$ is
not important; it scales the length unit in both directions, as can
be seen from Eq. (\ref{3.6}). Finally, we express $\alpha$ and $x$
by using $h(x)$ and $h'(x)$, $\alpha=h'(x)$ and $x=h(x)/h'(x)$, valid
for the linear cone, to place the results in the plot of $D(x)$ depending
on $h'(x)$ and $gh(x)$.

\begin{figure}
\includegraphics[scale=0.35]{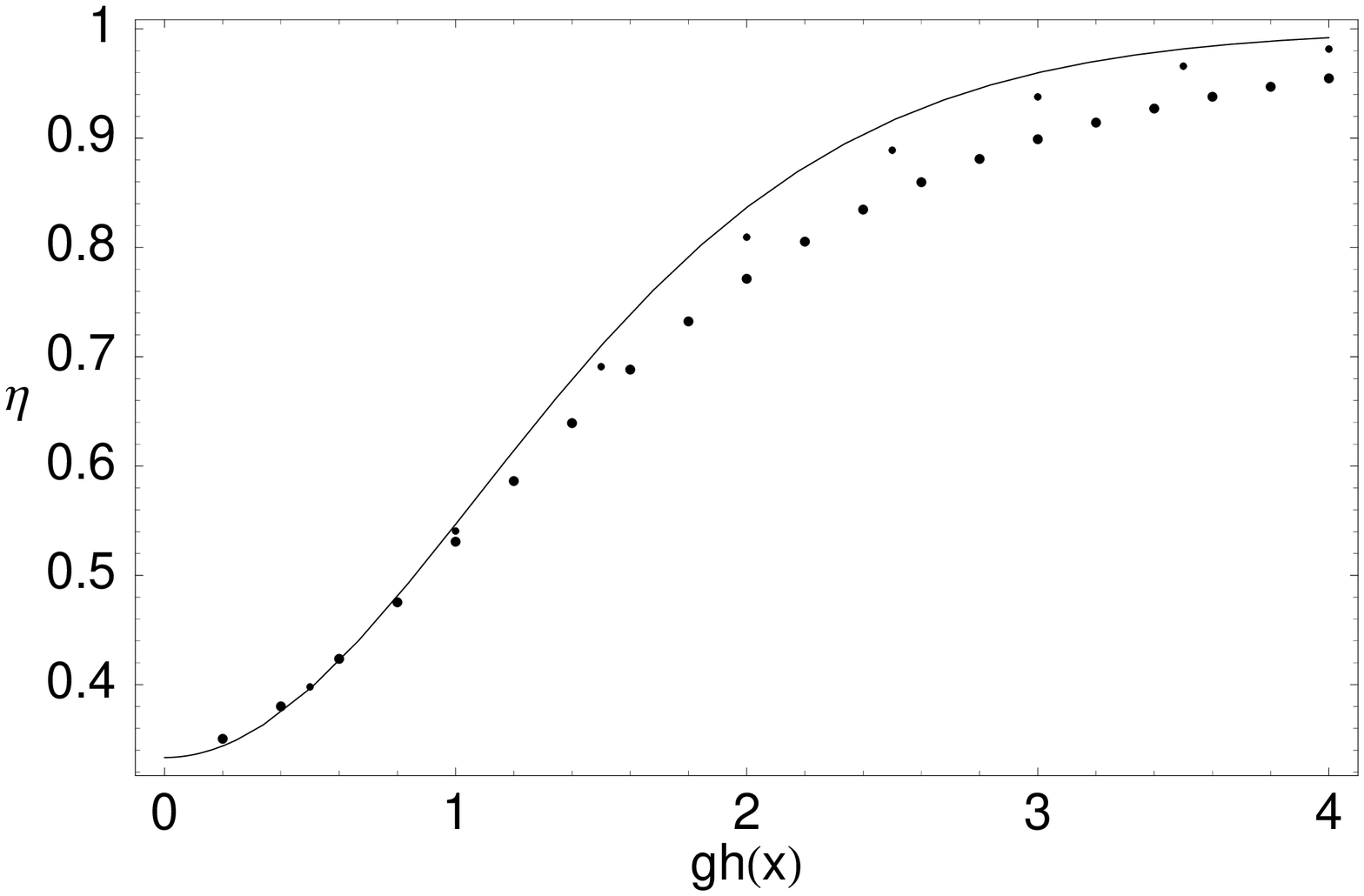}
\caption{The exponent $\eta$ plotted versus $gh(x)$ according to Eq.
(\ref{3.3}) (the line) and gained by fitting the interpolation formula
(\ref{3.2}) to $D(x)$ calculated for the linear cone, $\phi_0=\pi/6$
(the larger dots) and $\pi/10$ (the smaller upper dots)}.
\end{figure}
The data for $gh(x)=0.5$, $1$ and $2$ are depicted as dots in Fig. 2.
The interpolation formula describes the coefficient $D(x)$ satisfactorily
for small slopes of the boundaries $h'$ and close to the limits $gh(x)
\rightarrow 0$ and $\infty$. For larger $h'>0.5$ in an intermediate
region, roughly $1<gh(x)<5$, the deviations are more notable.
For practical purposes, one could try to find an interpolation formula
for $\eta$ fitting better the exact data obtained for the linear cones.
The exponents of Eq. (\ref{3.2}) fitted to the exact values of $D(x)$
for the cones with $\alpha=\tan(\pi/6)$ (the larger dots) and $\tan
(\pi/10)$ (the smaller dots) are depicted in Fig. 5 and compared with
the function (\ref{3.3}).

\section{IV. Conclusion}

The main aim of this paper was to arrive at an effective 1D description
of diffusion in a 2D symmetric channel of varying width $2h(x)$. The
diffusing particles are biased by a constant gravitational force $G$
acting in the direction perpendicular to the axis of the channel.

Our effective equation of the type (\ref{1.2}), governing evolution
of the 1D density $p(x,t)$ in the channel, goes beyond the Fick-Jacobs
approximation, considering only instant equilibration of the 2D density
in the transverse direction, which was used in the studies based on this
model \cite{SR,Das,Ray} till now. The effects of slower transverse 
relaxation are included in the
effective diffusion coefficient $D(x)$. We calculate this function
within a recurrence procedure \cite{map}-\cite{eff}, mapping rigorously
the 2D problem onto the longitudinal coordinate $x$ in the limit of
the stationary flow, i.e. when the net flux changes very slowly with
respect to the relaxation in the transverse direction.

The result is an expansion of $D(x)$ (\ref{2.22}) in a parameter
$\epsilon$ expressing the ratio of the diffusion constant in the
longitudinal and the transverse directions, $\epsilon=D_0/D_y$,
introduced artificially in the Smoluchowski equation (\ref{2.1})
and set to 1 at the end. Adding the transverse force makes the
result much more complicated, if compared with $D(x)$ (\ref{1.5})
for the diffusion alone. It is difficult to obtain a simple formula
usable in practice by direct summing of the expansion in $\epsilon
h'^2(x)$ up to infinity. So we suggest to use the interpolation
formula (\ref{3.2}) introduced before by Reguera and Rub\'i \cite{RR}
for diffusion. We showed that the biasing force effectively changes
the exponent $\eta$, depending on $gh(x)$, $g=G/k_BT$. It increases
from $1/3$ for negligible $G$ up to $1$ in strong fields. This dependence
can be approximated by the function (\ref{3.3}); then the first order
correction of the exact $D(x)$ (\ref{2.18}) is recovered.

The interpolation formula is compared with the truncated exact
expansion (\ref{2.22}) up to the 4-th order and also checked by the
model of biased diffusion in a linear cone, which is exactly solvable.
The agreement is satisfactory, but the differences increase especially
at steeper slopes of the boundaries, $h'(x)>0.5$, and intermediate
values of $gh(x)$ (roughly units). For practical purposes, one can try
to find some other (ad hoc) function for the exponent $\eta(gh(x))$ to
fit better the exact $D(x)$ in this region. Let us recall that the
"linear approximation" does not work well for $h'>1$ \cite{Berez}, because
higher derivatives play an important role there, too. Then taking more
complicated interpolation formulas depending only on $h'(x)$ may have
only a small effect on the further improvement of the results.

The calculation of the expansion of $D(x)$ (\ref{2.22}) presented
in the Section II also demonstrates how the mapping procedure
\cite{map,eff} can be applied to diffusion bounded in a channel
with hard walls and biased by a transverse force. Other possible
extensions are straightforward: we can add also a force acting along
the channel, or to go to 3D channels. The problem is growing complexity
of the expansions in $\epsilon$ and necessity of summation of at least
some group of their terms up to infinity; the expansions are converging
only in a restricted region of parameters, as seen in Fig. 2.
An effective way of searching for $D(x)$ is combination of the mapping
recurrence scheme with fitting the results of exactly solvable
models, as suggested in \cite{appr} and applied also in this work.

\section{Acknowledgments}
Support from VEGA grant No. 2/0113/11 and CE SAS QUTE project is
gratefully acknowledged.

\renewcommand{\theequation}{A\arabic{equation}}
\setcounter{equation}{0}

\section{Appendix A: Details of mapping}

We demonstrate here the mapping procedure on calculation of the first
order correction and then we prove the formula (\ref{3.1}) in the limit
of large $g$.

Starting from the zero-th order, we take $n=0$ in the recurrence
relation (\ref{2.14}), $\hat\omega_0=1$ and $\hat Z_0=-1$. We obtain
\begin{equation}\label{A1}
\partial_ye^{-gy}\partial_y\hat\omega_1=e^{-gy}\left(\frac{1}{A}\partial_x
A\partial_x-\partial_x^2\right)=e^{-gy}\frac{A'}{A}\partial_x\ .
\end{equation}
After applying Eq. (\ref{2.7}) and the first integration,
\begin{eqnarray}\label{A2}
\partial_y\hat\omega_1&=&e^{gy}\int dye^{-gy}gh'(x)
\coth{gh(x)}\partial_x\cr
&=& e^{gy}\left[-h'(x)\coth{gh(x)}e^{-gy}\partial_x+\hat C_1\right],
\end{eqnarray}
we fix the first integration constant $\hat C_1$ from the BC
(\ref{2.15}) at the upper boundary, $\partial_y\hat\omega_1=h'(x)
\partial_x$ at $y=h(x)$,
\begin{equation}\label{A3}
\hat C_1=h'(x)e^{-gh(x)}\left[1+\coth{gh(x)}\right]\partial_x.
\end{equation}
Notice that $\hat C_1$ is an operator, but independent of $y$.
Also one can check that the relation
\begin{equation}\label{A4}
\partial_y\hat\omega_1=\frac{h'(x)}{\sinh{gh(x)}}\left[e^{gy}
-\cosh{gh(x)}\right]\partial_x
\end{equation}
satisfies the BC (\ref{2.15}) at the lower boundary, $y=-h(x)$, too.
The next step is integration of Eq. (\ref{A4}),
\begin{equation}\label{A5}
\hat\omega_1=\frac{h'(x)}{\sinh{gh(x)}}\Big[\frac{1}{g}
e^{gy}-y\cosh{gh(x)}\Big]\partial_x+\hat C_0,
\end{equation}
and fixing the second integration constant $\hat C_0$ (again an operator)
from the normalization condition (\ref{2.16}). After some algebra, we
arrive at the formula (\ref{2.17}). Finally, applying Eq. (\ref{2.12})
to the resultant $\hat\omega_1$ gives the first order correction operator
$\hat Z_1$ (\ref{2.18}).

In the limit $g\rightarrow\infty$, we keep only the leading terms of
any expression during the calculation; the other terms, proportional
to powers of $e^{-gh(x)}$, are negligible. For $A(x)\simeq(1/g)e^{gh(x)}$,
the initial equation (\ref{A1}) of the recurrence scheme becomes
\begin{equation}\label{A6}
\partial_ye^{-gy}\partial_y\hat\omega_1=gh'(x)e^{-gy}\partial_x\ .
\end{equation}
After the first integration and fixing $\hat C_1$ at the upper boundary,
we get
\begin{equation}\label{A7}
\partial_y\hat\omega_1=h'(x)\left(2e^{g(y-h(x))}-1\right)\partial_x\ ;
\end{equation}
BC are satisfied at the lower boundary, too, because the term $\sim
e^{-2gh(x)}$ is negligible. The second integration and fitting the
normalization condition gives
\begin{equation}\label{A8}
\hat\omega_1=h'(x)\left[\frac{2}{g}e^{g(y-h(x))}-y+\frac{1}{g}-h(x)
\right]\partial_x
\end{equation}
up to the terms $\sim e^{-gh(x)}$ and smaller. Finally, in the
integration of $\hat Z_1$ according to Eq. (\ref{2.12}), only one term
remains,
\begin{equation}\label{A9}
\hat Z_1=ge^{-gh(x)}\int_{-h(x)}^{h(x)}dye^{-gy}h'^2(x)\simeq h'^2,
\end{equation}
nonvanishing in the limit of large $g$. Notice also that the exponential
term in Eqs. (\ref{A7}) and (\ref{A8}) does not contribute to
$\hat C_0$ and $\hat Z_1$; we can neglect it. This simplification
corresponds to fixing $\hat C_1$ at the lower boundary, then
$\partial_y\hat\omega_1=-h'(x)\partial_x$. The BC at the upper boundary
is not satisfied, but on the other hand, there are no particles there
for large $g$. We treat the upper boundary like it was in infinity.

Now we can prove the formula (\ref{3.1}). First we simplify the equation
(\ref{2.21}),
\begin{eqnarray}\label{A10}
\frac{1}{D(x)}&=&A(x)\left[1+\epsilon\hat Z+\epsilon^2\hat Z^2+...
\right]\frac{1}{A(x)}\cr
&\simeq& 1+\epsilon A(x)\hat Z\frac{1}{A(x)}+\left[\epsilon A(x)
\hat Z\frac{1}{A(x)}\right]^2+...,\hskip0.33in
\end{eqnarray}
hence
\begin{equation}\label{A11}
D(x)\simeq 1-\epsilon A(x)\hat Z\frac{1}{A(x)}=
1-\sum_{n=1}^{\infty}\epsilon^nA(x)\hat Z_n\frac{1}{A(x)};
\end{equation}
the difference depends on derivatives higher than $h'(x)$ and they are
neglected in our "linear" approximation.

The terms of the series in Eq. (\ref{A11}) can be expressed directly
by applying the operators $\hat\omega_n$ on a function $f(x)=\int dx
/A(x)\simeq \int ge^{-gh(x)}dx$. Then
\begin{equation}\label{A12}
A(x)\hat Z_n\frac{1}{A(x)}=-\int_{-h(x)}^{h(x)}dye^{-gy}\partial_x
\hat\omega_nf(x)
\end{equation}
from the relation (\ref{2.12}). The functions $\hat\omega_nf(x)$ are
derived by the same recurrence procedure, as it was demonstrated on
the operator $\hat\omega_1$ above. If we retain only the leading terms
in the limit $g\rightarrow\infty$ in our expressions and neglect
$h''(x)$ and its derivatives, we arrive at
\begin{eqnarray}\label{A13}
\hat\omega_1f(x)&=&-h'(x)e^{-gh(x)}\left[g\left(y+h(x)\right)-1
\right],\cr
\hat\omega_2f(x)&=&-\frac{h'^3}{2}e^{-gh}\left[g^2(y+h)^2
-4g(y+h)+2\right],\cr
\hat\omega_3f(x)&=&-\frac{h'^5}{6}e^{-gh}\big[g^3(y+h)^3-9g^2(y+h)^2\cr
&&\hskip0.7in +18g(y+h)-6\big],\cr
&...&\cr
\hat\omega_nf(x)&=&-{h'^{2n-1}}e^{-gh}\sum_{k=0}^n\frac{(-1)^{n-k}}
{k!}\left({n \above 0pt k}\right)\left[g(y+h)\right]^k.\cr &&
\end{eqnarray}
We can check normalization (\ref{2.16}) of these formulas,
\begin{eqnarray}\label{A14}
\int_{-h}^hdy e^{-gy-gh}\sum_{k=0}^{n}\frac{(-1)^{n-k}}{k!}
\left({n \above 0pt k}\right)\left[g(y+h)\right]^k\cr\simeq
\sum_{k=0}^{n}\frac{(-1)^{n-k}}{g\ k!}\left({n \above 0pt k}\right)
\int_0^{\infty}e^{-z}z^kdz=0,
\end{eqnarray}
after substituting $z=g(y+h)$ and replacing the upper boundary $2gh$
by infinity; we omitted writing explicit dependence of $h(x)$ on $x$.
The coefficients of the expansion (\ref{A11}) are integrated in a
similar way; after completing the $x$ derivative of the formulas
(\ref{A13}) in Eq. (\ref{A12}) and using the normalization (\ref{A14}),
only the term
\begin{eqnarray}\label{A15}
A(x)\hat Z_n\frac{1}{A(x)}=\hskip2.2in\cr
\int_{-h}^hgdye^{-gy-gh}h'^{2n}
\sum_{k=1}^n\frac{(-1)^{n-k}}{(k-1)!}\left({n \above 0pt k}\right)
[g(y+h)]^{k-1}\cr
=h'^{2n}\sum_{k=1}^n(-1)^{n-k}\left({n \above 0pt k}\right)
=-(-1)^nh'^{2n}\hskip0.3in
\end{eqnarray}
remains, the result we wanted to prove.

Finally, one can check by direct calculation that the formulas (\ref{A13})
satisfy the recurrence relation (\ref{2.14}) and the BC (\ref{2.15}) at the
lower boundary. The operators $\hat\omega_n$ are replaced here by the
functions $\hat\omega_nf(x)$, $\partial_xf(x)=ge^{-gh(x)}$. The terms
depending on $\hat Z_k$ disappear from Eq. (\ref{2.14}), since
$A(x)\hat Z_k(1/A(x))=(-1)^{(n+1)}h'^{2n}$ according to Eq. (\ref{A15}),
and its derivative depends on $h''$, which is neglected.

\renewcommand{\theequation}{B\arabic{equation}}
\setcounter{equation}{0}

\section{Appendix B: Exact solution}

We present here the stationary solution of the Smoluchowski equation
(\ref{2.1}) for the biased diffusion in a linear cone.

For a point-like source of particles placed at the origin of the
coordinate system, the stationary equation (\ref{2.1}),
\begin{equation}\label{B1}
0=\partial_x^2\rho(x,y)+\partial_y e^{-gy}\partial_y e^{gy}\rho(x,y)
\end{equation}
becomes separable after substitution
\begin{equation}\label{B2}
\rho(x,y)=e^{-gy/2}u(x,y)
\end{equation}
and converting to the polar coordinates, $x=r\cos{\phi}$, $y=r\sin{\phi}$,
\begin{equation}\label{B3}
\left[\frac{1}{r}\partial_r r\partial_r+\frac{1}{r^2}\partial_{\phi}^2
-\left(\frac{g}{2}\right)^2\right]u(r,\phi)=0.
\end{equation}
The particular solutions are $u(r,\phi)=R_n(gr/2)e^{in\phi}$; $R_n$
stands for the Bessel $I_n$ or the Bessel $K_n$ functions. In an
unbounded plane, the linear combinations describing the biased
diffusion cannot contain the $I_n$ functions, since $\rho(x,y)$
would diverge for $y\rightarrow -\infty$. The solution $u(x,y)$ is
then composed from the Bessel $K_n$ functions. The "ground-state"
solution $u_0(r,\phi)=K_0(gr/2)$ carries the flux along the force.
Other particular solutions $u_n(r,\phi)=K_n(gr/2)\sin(n\phi)$
are necessary for fitting $\rho(x,y)=0$ at an absorbing boundary,
if it is considered, but they do not contribute to the net flux; one
can check that
\begin{equation}\label{B4}
\int_{-\infty}^{\infty}e^{-gy}\partial_y\left[u_n(x,y)e^{gy/2}\right]dx
=-J\delta_{n,0}
\end{equation}
at any fixed $y<0$ (below the source). We need only the ground
state $u_0(x,y)$ for the calculation of $D(x)$, Eq. (\ref{3.5}).

Our problem is to find such a solution for diffusion in a linear
cone, i.e. satisfying the BC (\ref{2.3}) at $y=\pm\alpha x$. In
the polar coordinates, the BC become
\begin{equation}\label{B5}
\partial_{\phi}u(r,\phi)=-\frac{gr}{2}\cos(\phi)u(r,\phi)
\Big\vert_{\phi=\pm\phi_0},
\end{equation}
for any $r>0$ and $h'(x)=\alpha=\tan{\phi_0}$.

\begin{figure}
\includegraphics[scale=0.3]{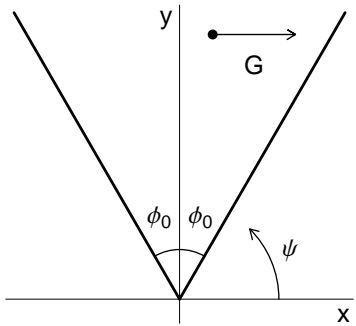}
\caption{The linear cone in Fig. 3 rotated by $\pi/2$}
\end{figure}
This task is related to the calculation of the 2D stationary density
of particles dragged out of the cone by a constant force along the
$x$ axis \cite{forced}. If we rotate our channel in Fig. 3 by $\pi/2$,
we get the relevant picture, Fig. 6. In comparison to the previous
problem, the particles diffuse in a different sector; instead of
the angle $\psi=\phi+\pi/2\in(0,\pi/2-\phi_0),$ they are confined
in $\psi\in(\pi/2-\phi_0,\pi/2+\phi_0)$. For certain values of
$\phi_0$, we can extend the known solutions in the sector adjacent
to the $x$ axis \cite{forced} to the sector of our interest.

We recall briefly the stationary solution of the Smoluchowski equation
in the sector $\psi\in(0,\psi_0)$. After rotation of the coordinate system,
$\phi$ is simply replaced by $\psi$ in Eq. (\ref{B3}) and the rotated
BC (\ref{B5}),
\begin{equation}\label{B6}
\partial_{\psi}u(r,\psi)=-\frac{gr}{2}\sin(\psi)u(r,\psi)
\end{equation}
has to be satisfied at $\psi=\pm\psi_0$. To express the solutions
$u(r,\psi)$ here, we are inspired by the integral representation
of the Bessel functions
$K_{\nu}$ \cite{RG},
\begin{equation}\label{B7}
K_{\nu}(r)=\int_0^{\infty}e^{-r\cosh{t}}\cosh{\nu t}\ dt\ .
\end{equation}
One can check by direct calculation, that the integral
\begin{equation}\label{B8}
u(r,\psi)=\int_0^{\infty}e^{-(gr/2)\cosh{t}}\Big[f(t+i\psi)
+f(t-i\psi)\Big]dt\ ,
\end{equation}
is a solution of the equation (\ref{B3}) (with $\phi$ replaced by
$\psi)$ for any even analytic function $f(z)=f(-z)$ of the complex
variable $z$ having no pole along the integration path. Notice also
that the function (\ref{B8}) has expected symmetry $u(r,\psi)=
u(r,-\psi)$, given by the direction of the force along the $x$ axis.

The function $f(z)$ is fixed from the BC (\ref{B6}) at $\psi=\pm\psi_0.$
Applying the formula (\ref{B8}) in Eq. (\ref{B6}) and integrating by
parts, we obtain the condition
\begin{eqnarray}\label{B9}
\big[f(t+i\psi_0)-f(t-i\psi_0)\big]\sinh{t}=\hskip1in\cr
=i\big[f(t+i\psi_0)+f(t-i\psi_0)\big]\sin{\psi_0} \ \ 
\end{eqnarray}
valid for any $t\ge 0$. If we write $f(z)=g(z)\tanh{(z/2)}$,
$g(z)$ has to satisfy $g(t+i\psi_0)=g(t-i\psi_0)$.
Then the "ground state" solution $u_0(r,\psi)$ is generated by
$g_0(z)=\coth(\pi z/2\psi_0)$ and the other particular solutions
$u_n(r,\psi)$ come from $g_n(z)=\sinh(n\pi z/\psi_0)$.
Again, only $u_0(r,\psi)$ is connected with the 1D stationary flux
flowing along the $x$ axis, and $u_{n>0}$ are modes projected out
by the mapping procedure \cite{forced}, which are not necessary for
calculation of $D(x)$.

To get the solution $u_0$ for the cone with the transverse field, i.e.
for the sector $\psi\in(\pi/2-\phi_0,\pi/2+\phi_0)$, we have to find
the function $f(z)$ such that the BC (\ref{B6}) are satisfied at
$\psi=\pi/2\pm\phi_0$. The same treatment leads to a condition similar
to Eq. (\ref{B9}); if we write $f(z)=g(z)\tanh(z/2)$, then 
$g(t+i\psi)=g(t-i\psi)$ is required at both boundaries,
$\psi=\pi/2\pm\phi_0$, and any $t\ge 0$.

For specific angles $\phi_0$, we can adopt the function
\begin{equation}\label{B10}
g_0(z)=\coth(mz/2)=\frac{e^{mz}+1}{e^{mz}-1}.
\end{equation}
It satisfies the required condition not only at $\psi_0=\pi/m$, used in
the sector adjacent to the $x$ axis, but also at any its integer multiple.
To get the "ground state", we need to adjust $\pi/2\pm\phi_0$ to be
succeeding integer multiples of $\pi/m$; the imaginary part of $m(t+i\psi)$
has to change by $i\pi$ if $\psi$ increases from $\pi/2-\phi_0$ up to
$\pi/2+\phi_0$. These requirements are met for odd numbers $m\ge 3$.
For the corresponding angles $\phi_0=\pi/2m=\pi/6,\pi/10,..$, the formula
(\ref{B10}) becomes the function $g_0(z)$ generating the "ground
state" $u_0(r,\psi)$ also in the sector of our interest.

Still, there is a problem at the boundary whose angle $\psi$ is an
even multiple of $\psi_0$; the function $g_0(z)$ and also the
corresponding $f(z)$ have a pole at $t=0$. We solve it by
changing the integration path in the complex plane.

First we return back to the unrotated coordinate system and the angle
$\phi$. The variable $t$ in the integral (\ref{B8}),
\begin{eqnarray}\label{B11}
u(r,\phi)&=&\int_0^{\infty}e^{-(gr/2)\cosh{t}}\Big[f(t+i\phi+i\pi/2)\cr
&&\hskip0.5in+f(t-i\phi-i\pi/2)\Big]dt\ ,
\end{eqnarray}
can be substituted by $t=z\pm i\pi/2$ and the path is then shifted in
the complex plane by $\mp i\pi/2$ correspondingly. In the final
formula, rewritten in a symmetric way,
\begin{eqnarray}\label{B12}
2u(r,\phi)&=&\int_0^{i\pi/2+\infty}e^{-(gr/2)\cosh(z-i\pi/2)}\Big[
f(z+i\phi)\cr &&\hskip0.3in +f(z-i\phi-i\pi)\Big]dz\cr
&+&\int_0^{-i\pi/2+\infty}e^{-(gr/2)\cosh(z+i\pi/2)}\Big[
f(z-i\phi)\cr &&\hskip0.3in +f(z+i\phi+i\pi)\Big]dz,
\end{eqnarray}
we change the lower limits $\pm i\pi/2$ by zero and the integration
path in both integrals avoids the poles at $z=\pm i\pi/2$ from the
right, see Fig. 7.
\begin{figure}
\includegraphics[scale=0.33]{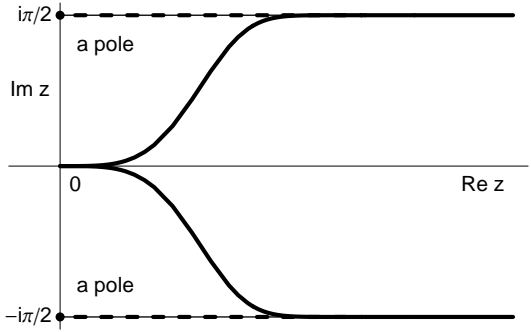}
\caption{Integration paths in Eq. (\ref{B12}). The paths crossing
poles on the imaginary axis (the dashed lines) are replaced by the
full lines, avoiding the poles from the right.}
\end{figure}

Notice that the function (\ref{B12}) holds the symmetry $u(r,\phi)=
u(r,-\phi-\pi)$, determined now by the $y$ direction of the force
(Fig. 3). Direct calculation shows that also this $u(r,\phi)$
with changed lower limits of integration solves the equation
(\ref{B3}) for any $f(z)=f(-z)$, which has no pole along the
integration path. Substituting Eq. (\ref{B12}) in the BC (\ref{B5})
and integrating by parts results in the conditions $g(z\pm i\phi)=
g(z\mp i\phi\mp i\pi)$ at $\phi=\pm\phi_0$ and any $z$ on the integration
path; we rewrote again $f(z)=g(z)\tanh(z/2)$. It is easy to verify
that the function (\ref{B10}) satisfies these conditions for
$\phi_0=\pi/2m$, $m=3,5,...$, so taking
\begin{equation}\label{B13}
f(z)=\coth(mz/2)\tanh(z/2)
\end{equation}
in Eq. (\ref{B12}), we obtain the "ground state" solution
$u_0(r,\phi)$ in the linear cone with a transverse force for these
specific slopes $\alpha=\tan{\phi_0}$.

In the resultant stationary density, two integration constants can
be added,
\begin{equation}\label{B14}
\rho(x,y)=C_1e^{-gy/2}u_0(x,y)+C_0e^{-gy}.
\end{equation}
$C_1$ controls the stationary net flux connected with the density
$\rho(x,y)$ and $C_0$ sets the BC $\rho(x,y)=0$ at a distant boundary
absorbing the particles. None of them influences the calculation of
$D(x)$. The contribution of the term proportional to $C_0$ to
$p(x)/A(x)$ is constant and so it gives zero in the formula (\ref{3.5}).
The net flux $J$, as well as $\partial_x[p(x)/A(x)]$, are proportional to
$C_1$ and so it is canceled in the resultant $D(x)$.

\end{document}